\documentclass[onecollarge,natbib]{svjour2}
\bibpunct{[}{]}{;}{n}{}{,} 
\smartqed  
\usepackage{graphicx}
\usepackage{amssymb,amsmath}

\journalname{Archive of Applied Mechanics}
\begin{document}

\title{Baryon spectroscopy with polarization observables from CLAS%
  \thanks{Supported in parts by the U.S. National Science Foundation: NSF PHY-1505615.}
}


\author{Steffen Strauch for the CLAS Collaboration}


\institute{S. Strauch \at
              University of South Carolina \\
              Tel.: +1-803-777-8197\\
              Fax: +1-803-777-3065\\
              \email{strauch@sc.edu}           
}

\date{Received: date / Accepted: date}

\maketitle

\begin{abstract}
  The spectrum of nucleon excitations is dominated by broad and
  overlapping resonances.  Polarization observables in photoproduction
  reactions are key in the study of these excitations.  They give
  indispensable constraints to partial-wave analyses and help clarify
  the spectrum.  A series of polarized photoproduction experiments
  have been performed at the Thomas Jefferson National Accelerator
  Facility with the CEBAF Large Acceptance Spectrometer (CLAS).  These
  measurements include data with linearly and circularly polarized
  tagged-photon beams, longitudinally and transversely polarized
  proton and deuterium targets, and recoil polarizations through the
  observation of the weak decay of hyperons.  An overview of these
  studies and recent results will be given.
\end{abstract}

\section{Introduction}

The nucleon is a color-neutral object which consists of color-charged
quarks and gluons.  Quantum chromodynamics (QCD) is the fundamental
theory of the strong interaction between quarks and gluons.  Valuable
information about QCD can be learned from nuclear spectroscopy;
\textit{e.g.}  information about the internal degrees of freedom in a
nucleon.  Experiments, especially pion-nucleon scattering, have
confirmed low-lying excited states predicted by quark models with
three independent quark degrees of freedom.  These models, however,
predict an overabundance of higher-lying excited states compared to
what has been observed until now \cite{Crede:2013kia}.  Also, recent
lattice QCD calculations \cite{Edwards:2011jj} find a large number of
not-yet-discovered nucleon resonances.  To clarify the nucleon
resonance spectrum is an important task in the study of QCD.

Nucleon resonances are short-lived, and the identification of
resonances in partial-wave analyses of experimental data is
complicated by their large width and overlap.  Experimental cross
sections are insufficient, and polarization observables are crucial to
constrain these analyses.  A complete set of certain polarization
observables is necessary to unambiguously determine the amplitudes of
the reaction.  In the photoproduction of pseudoscalar mesons, a
formally complete experiment requires at each energy and angle the
measurement of at least eight carefully chosen observables
\cite{Chiang:1996em}.  In the photoproduction of two mesons, even more
observables are needed \cite{Roberts:2005mn}.  It is also important to
include in the analysis data from a variety of excitation and decay
channels, as some of the missing states may couple only weakly to,
\textit{e.g.}, the $\pi N$ channel.

In the following, examples of recent photoproduction measurements of
polarization observables from the CLAS Collaboration are presented. A
similar overview has been given in \cite{Strauch:2015noa}.  These
measurements include single pseudoscalar-meson, vector-meson,
double-pion, and hyperon photoproduction off the proton and
quasi-free off the bound neutron.  The observables include single-
and double-polarization observables with combinations of polarized
beam, target, and the polarization of the recoiling baryon.

The experiments were performed in Experimental Hall B at the Thomas
Jefferson National Accelerator Facility (JLab).  The incident
bremsstrahlung photon beams were energy-tagged \cite{Sober:2000we} and
either unpolarized, circularly, or linearly polarized. The photon beam
irradiated the production target.  Unpolarized liquid hydrogen and
deuterium targets, as well as the newly developed polarized frozen-spin (FROST)
\cite{Keith201227} and HDice targets \cite{Sandorfi:2013gra,PR06101},
have been used in these experiments.  Final-state particles were
detected in the CEBAF Large Acceptance Spectrometer (CLAS)
\cite{Mecking:2003zu}.  Recoil polarization was accessable in
hyperon-production measurements through the measurement of the
decay-proton angular distribution in the parity-violating weak decay
of hyperons.

\section{Unpolarized Targets}

The CLAS Collaboration has studied the photoproduction reactions
$\gamma p \to p\pi^0$ and $\gamma p \to n\pi^+$ with
linearly-polarized photons in an energy range from 1.10 to 1.86~GeV
\cite{Dugger:2013crn}.  The beam asymmetry observable $\Sigma$ has
been obtained from the pion angular distributions with respect to the
polarization direction of the linearly-polarized photon beam.  The
high statistics in these reactions allowed for precise constraints of
partial-wave analyses. Resonance couplings have been extracted in fits
of the SAID \cite{Workman:2012jf} analysis after including the new
data set.  The largest change from previous fits was found to occur
for the 'well known' $\Delta(1700)3/2^-$ and $\Delta(1905)5/2^+$
resonances \cite{Dugger:2013crn}.

\begin{figure}[!htb]
\centering
\includegraphics[width=\textwidth]{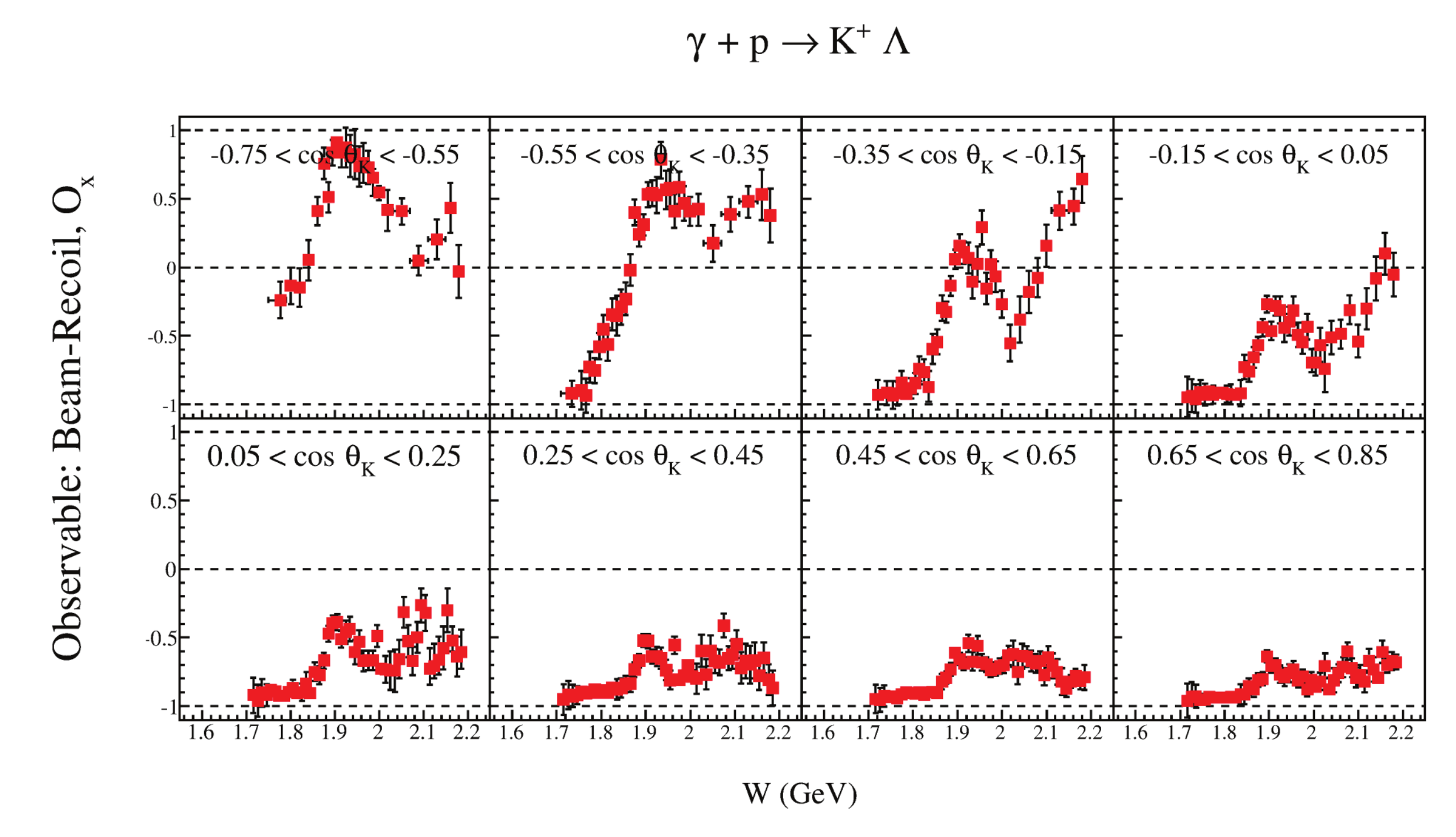}
\caption{Preliminary energy distributions of CLAS
  $\gamma p \to K^+ \Lambda$ data for the double-polarization
  observable $O_x$ in various kaon angular bins.  Figure from D.~Ireland
  (University of Glasgow).}
\label{fig1}
\end{figure}

Polarization observables were also obtained in the
hyperon-photoproduction reactions $\gamma p \to K^+\Lambda$ and
$\gamma p \to K^+\Sigma^0$.  Results include the $\Lambda$ recoil
polarization $P$ from measurements with unpolarized photon beams
\cite{McNabb:2003nf,McCracken:2009ra} as well as the
polarization-transfer observables $C_x$ and $C_z$ with
circularly-polarized beams \cite{Bradford:2006ba}.  These data were
critical in a coupled-channel analysis by the Bonn-Gatchina group
\cite{Sarantsev:2005tg,Nikonov:2007br}.  In particular, the analysis
found further evidence for the, at the time, poorly known
$N(1900)3/2^+$ resonance.  This resonance is predicted by symmetric
three-quark models, but is not expected to exist in earlier
quark-diquark models.  Preliminary hyperon-photoproduction data with
linearly-polarized photons off an unpolarized proton target have been
obtained up to $W \approx 2.2$~GeV.  Together with the recoil
polarization of the hyperon, this gives access to five polarization
observables: $\Sigma$, $P$, $T$, $O_x$, and $O_z$.  Energy
distributions of the preliminary results of the beam-recoil
polarization observable $O_x$ for the $ K^+ \Lambda$ channel are shown
in Fig.~\ref{fig1}.

\section{Polarized Targets}

An integral part of the experimental N* program at JLab are
experiments off polarized protons
\cite{PR02112,PR03105,PR04102,PR05012,PR06013}.  The Jefferson Lab
frozen spin target (FROST) was constructed for use inside CLAS to
allow for the measurement of polarization observables with
longitudinally- or transversely-polarized protons in the butanol target
material \cite{Keith201227}.

\begin{figure}[!htb]
\centering
\includegraphics[width=\textwidth]{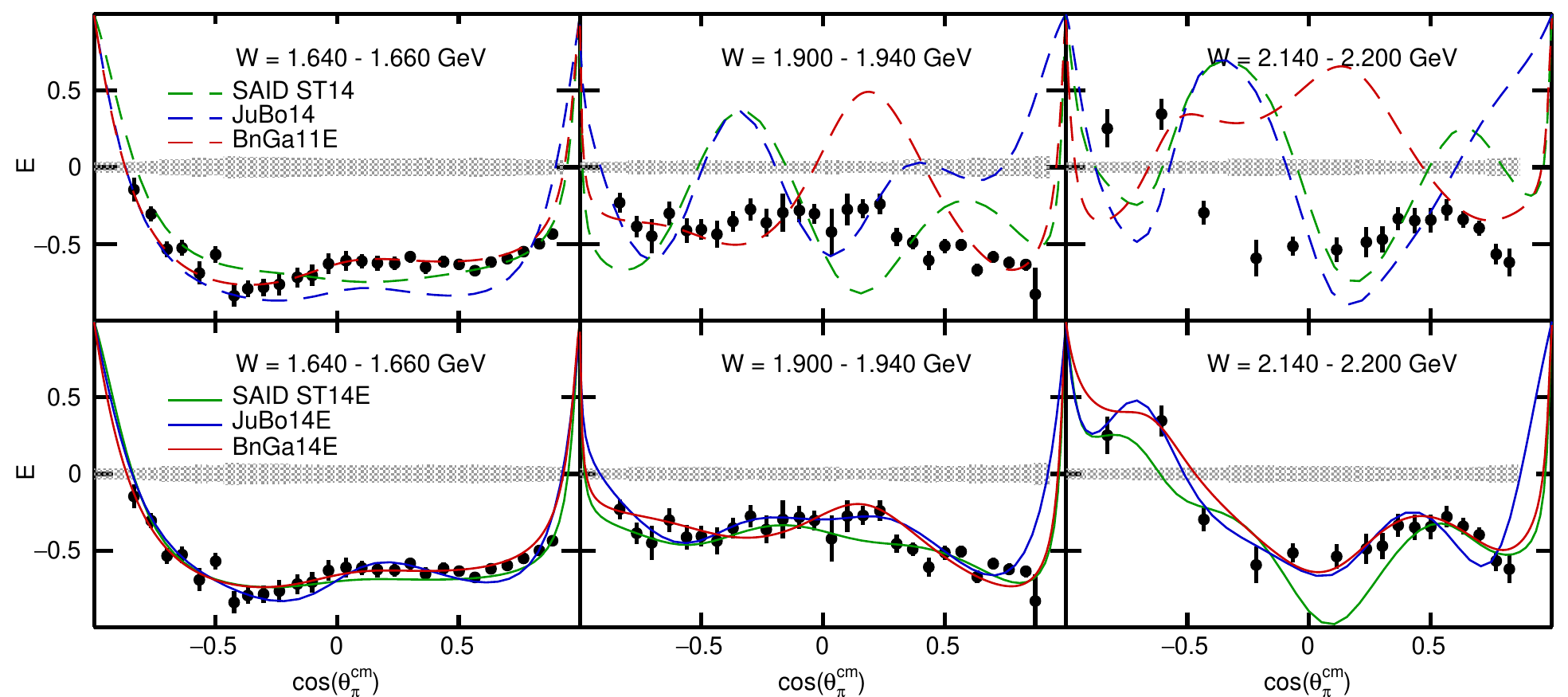}
\caption{CLAS data for the beam-target polarization observable $E$ in
  the $\gamma p \to \pi^+n$ reaction for selected center-of-mass
  energy bins.  The shaded bands indicate systematic uncertainties.
  The dashed curves in the upper panels are results from the SAID ST14
  \cite{Workman:2012jf}, J\"ulich14 \cite{Ronchen:2014cna}, and
  BnGa11E \cite{Anisovich:2011fc} analyses.  The solid curves in the
  lower panels are results from updated analyses including the new
  CLAS $E$ data. Figure from Ref.~\cite{Strauch:2015zob}.}
\label{fig2}
\end{figure}

First results from the FROST program have been published for the
$\gamma p \to \pi^+n$ reaction \cite{Strauch:2015zob}.  The
double-polarization observable $E$ has been determined in the energy
range from 0.35 to 2.37~GeV from data of circularly-polarized photons
incident on longitudinally-polarized protons.  A subset of the about
900 data points is shown in the three energy bins of
Fig.~\ref{fig2}.  Results from previous
partial-wave analyses describe the new data at low photon energies
reasonably well; at higher energies, however, significant deviations are
observed.  The data have been included in new analyses resulting in
good descriptions of the data and in updated nucleon resonance
parameters.  One particularly interesting result is strengthened
evidence for the poorly known $\Delta (2200)7/2^-$ resonance in
improving the Bonn-Gatchina fit at the highest
energies~\cite{Anisovich:2015gia}.  The mass of the
$\Delta (2200)7/2^-$ resonance is significantly higher than the mass
of its parity partner $\Delta(1950)7/2^+$, which is the lowest-mass
$\Delta^*$ resonance with spin-parity $ J^P=7/2^+$.

\begin{figure}[!htb]
\centering
\includegraphics[width=0.7\textwidth]{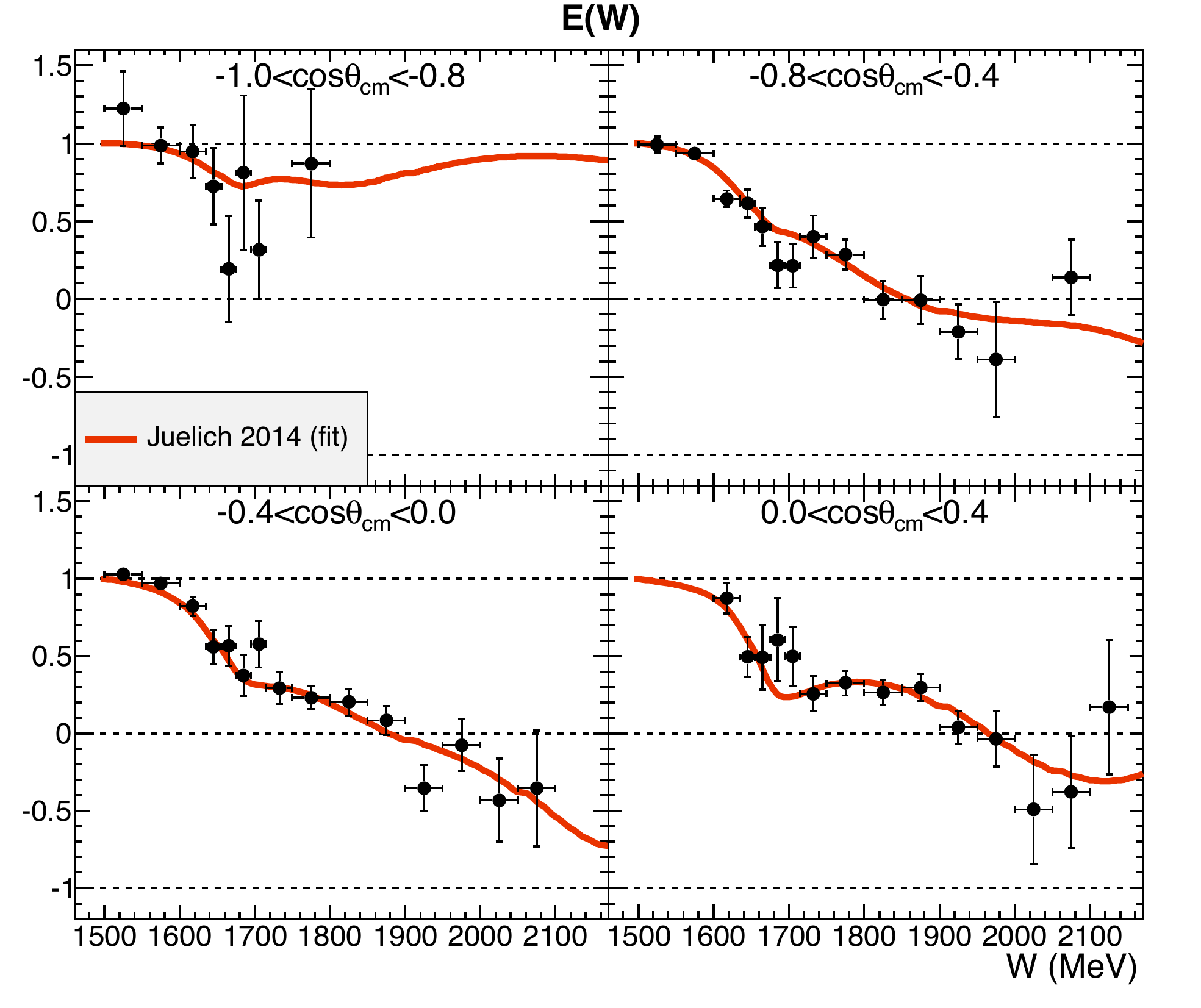}
\caption{Observable $E$ in the $\gamma p \to \eta p$ reaction as a
  function of $W$ for various angular bins.  The red curve shows
  fits with the J\"ulich-Bonn dynamical coupled-channel model.  Figure
  from Ref.~\cite{Senderovich:2015lek}.}
\label{fig3}
\end{figure}

The polarization observable $E$ has also been measured in the
$\gamma p \to \eta p$ reaction from threshold to $W = 2.15$~GeV
\cite{Senderovich:2015lek}. Because $\eta$ mesons have isospin zero,
the reaction selects isospin-$1/2$ resonances in the nucleon resonance
spectrum.  Figure \ref{fig3} shows the data and a
fit with the J\"ulich-Bonn dynamical coupled-channel model.  It has
been shown in Ref.~\cite{Senderovich:2015lek} that the observable $E$
in $\eta$ photoproduction is especially suited to disentangle
electromagnetic resonance properties.  Initial investigation of these
results show pronounced changes in the description of this observable
when these new CLAS data are included.  The fit in
Fig.~\ref{fig3} describes the data quite well
without the need for an additional narrow resonance near 1.68~GeV,
which was previously suggested \cite{Senderovich:2015lek}.

\begin{figure}[!htb]
\centering
\includegraphics[width=0.47\textwidth]{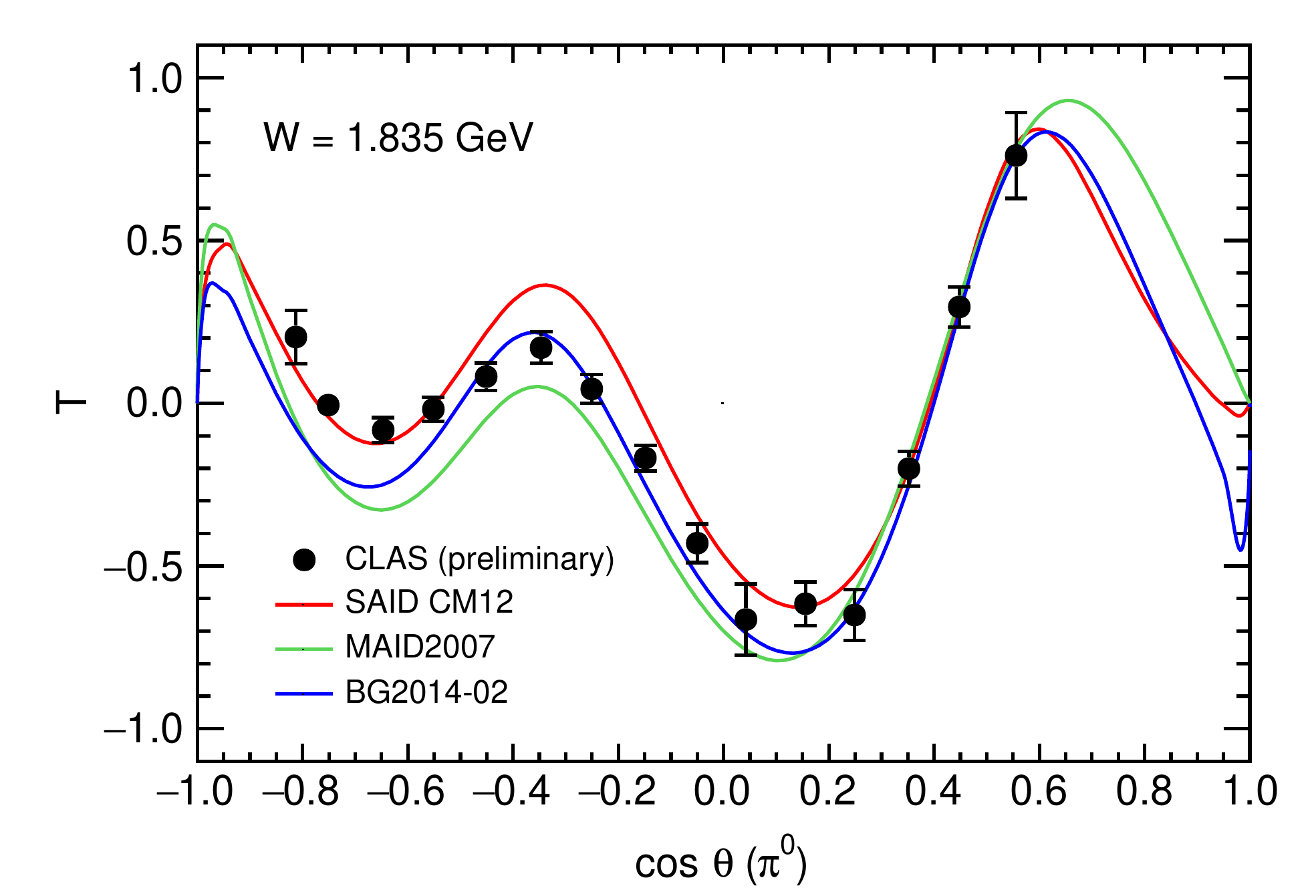}
\includegraphics[width=0.47\textwidth]{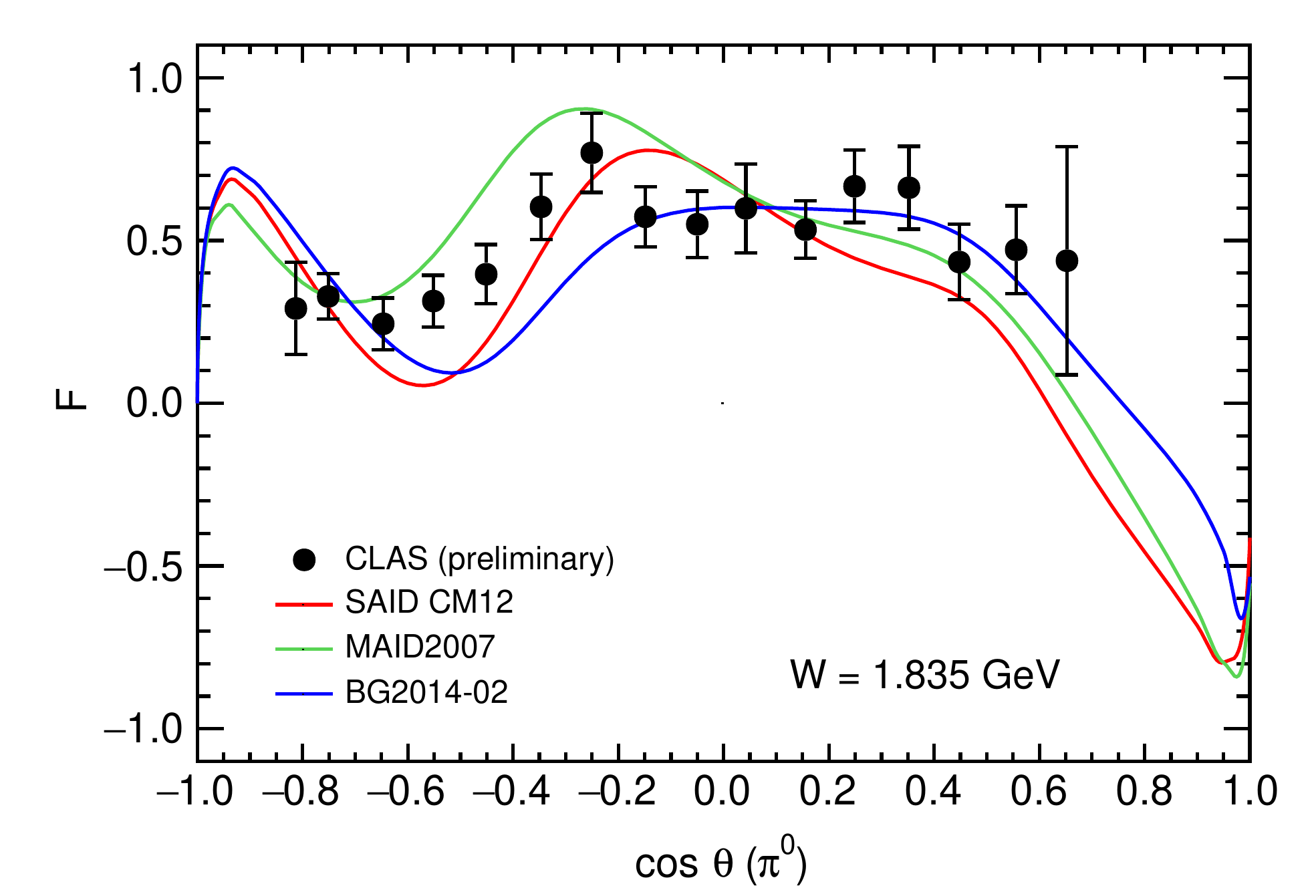}
\caption{Preliminary CLAS results for the observables $T$ and $F$ in
  the $\gamma p \to \pi^0 p$ reaction as a function of
  $\cos\theta_{\pi}$ for $W = 1.835$~GeV. The data are compared with
  various partial-wave analyses (not fitted to the data)
  from the SAID \cite{Workman:2012jf}, MAID \cite{Drechsel:2007if},
  and Bonn-Gatchina \cite{Anisovich:2011fc} groups. Figures from
  H.~Jiang (University of South Carolina).}
\label{fig4}
\end{figure}

FROST data from other single-pion photoproduction channels and for
other observables are under ongoing analyses.
Examples of preliminary angular distributions for the
target $T$ and beam-target $F$ polarization observables in the
reaction $\gamma p \to \pi^0 p$ are shown in
Fig.~\ref{fig4}.  These observables are extracted
from data with unpolarized and circularly-polarized photons off
transversally-polarized protons, respectively.  In the figure, the data are compared
with present results of partial-wave analyses which do not include
these preliminary data.

\begin{figure}[!htb]
\centering
\includegraphics[width=\textwidth]{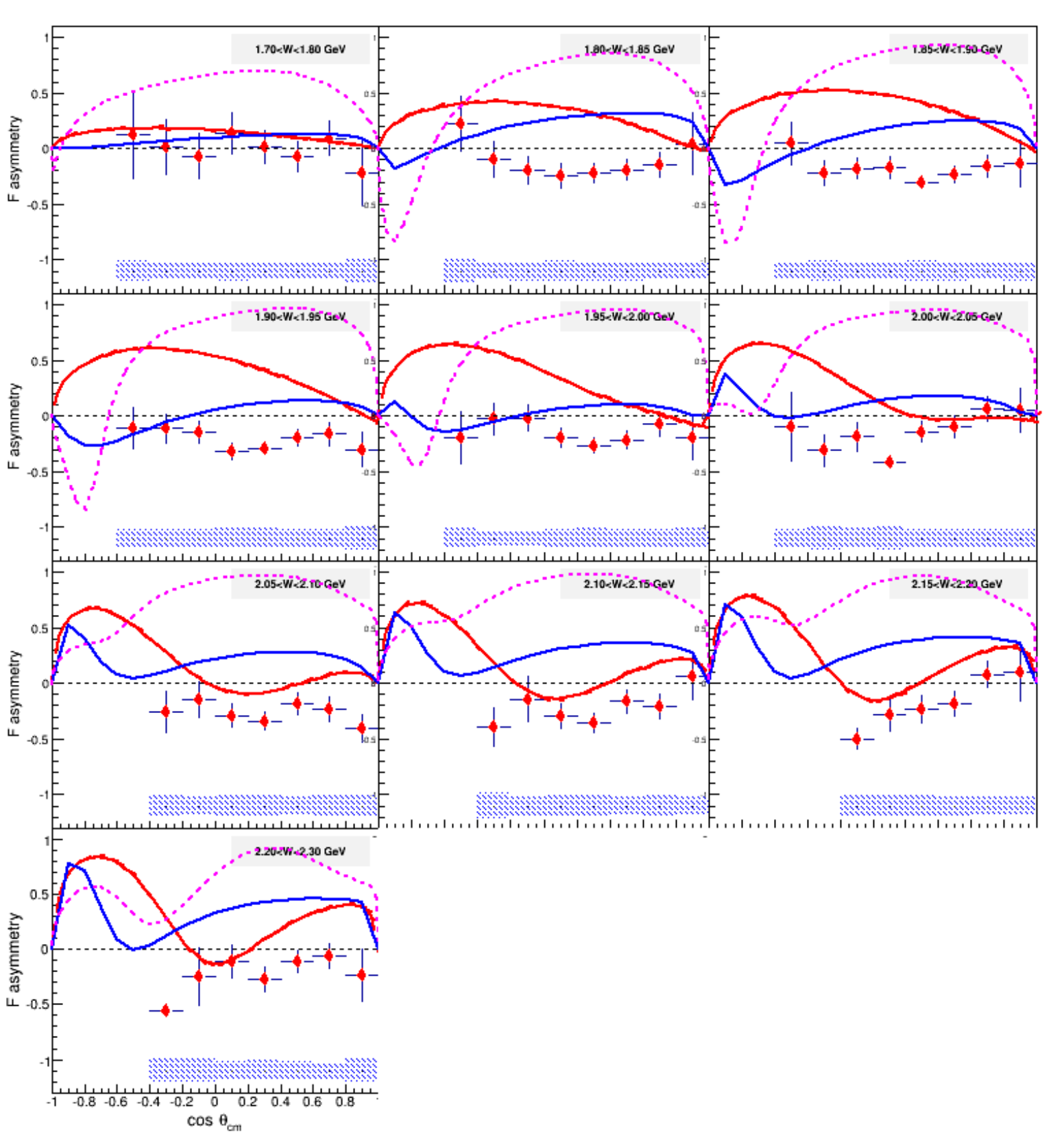}
\caption{Preliminary angular distributions of the polarization
  observable $F$ from FROST in the $\gamma p \to K^+\Sigma^0$ reaction
  for various $W$ bins.  Model curves are from the RPR-Ghent
  \cite{WalfordRPR} (red), KAON-MAID \cite{WalfordKM} (blue), and
  Bonn-Gatchina \cite{Anisovich:2011fc} (magenta) models.  Figure 
  from N.~Walford (U.~Basel).}
\label{fig5}
\end{figure}

Hyperon photoproduction reactions are also studied with data from
FROST.  Figure \ref{fig5} shows preliminary
results of the beam-target observable $F$.  The data cover energies
between $W = 1.7$~GeV and $W = 2.3$~GeV.  The data are compared to
prior results of analyses with the RPR-Ghent \cite{WalfordRPR},
KAON-MAID \cite{WalfordKM}, and Bonn-Gatchina \cite{Anisovich:2011fc}
models.  As none of the models describe the data well the FROST data
will provide important new constraints to the models.

\begin{figure}[!htb]
\centering
\includegraphics[width=0.45\textwidth]{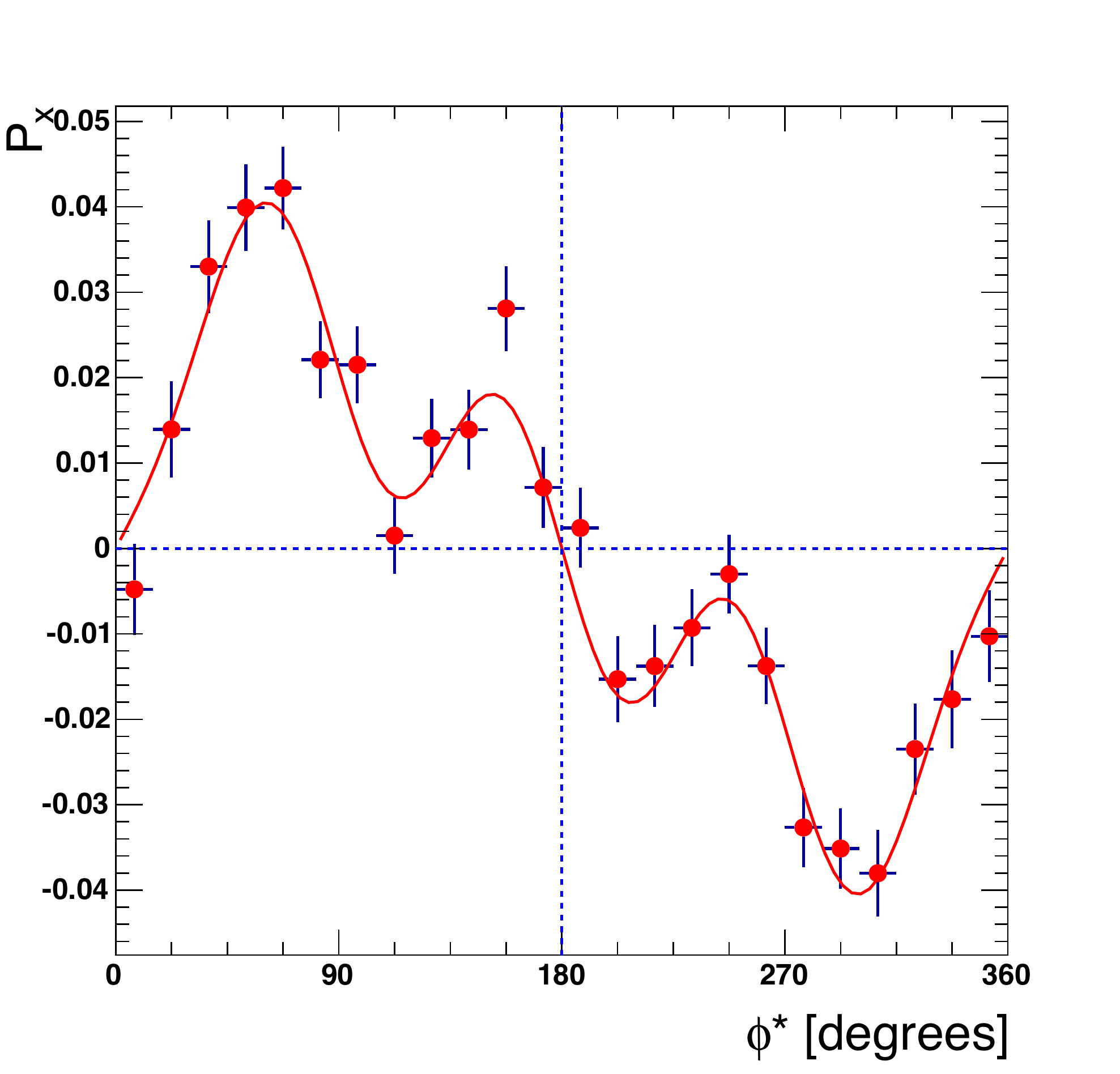}%
\includegraphics[width=0.45\textwidth]{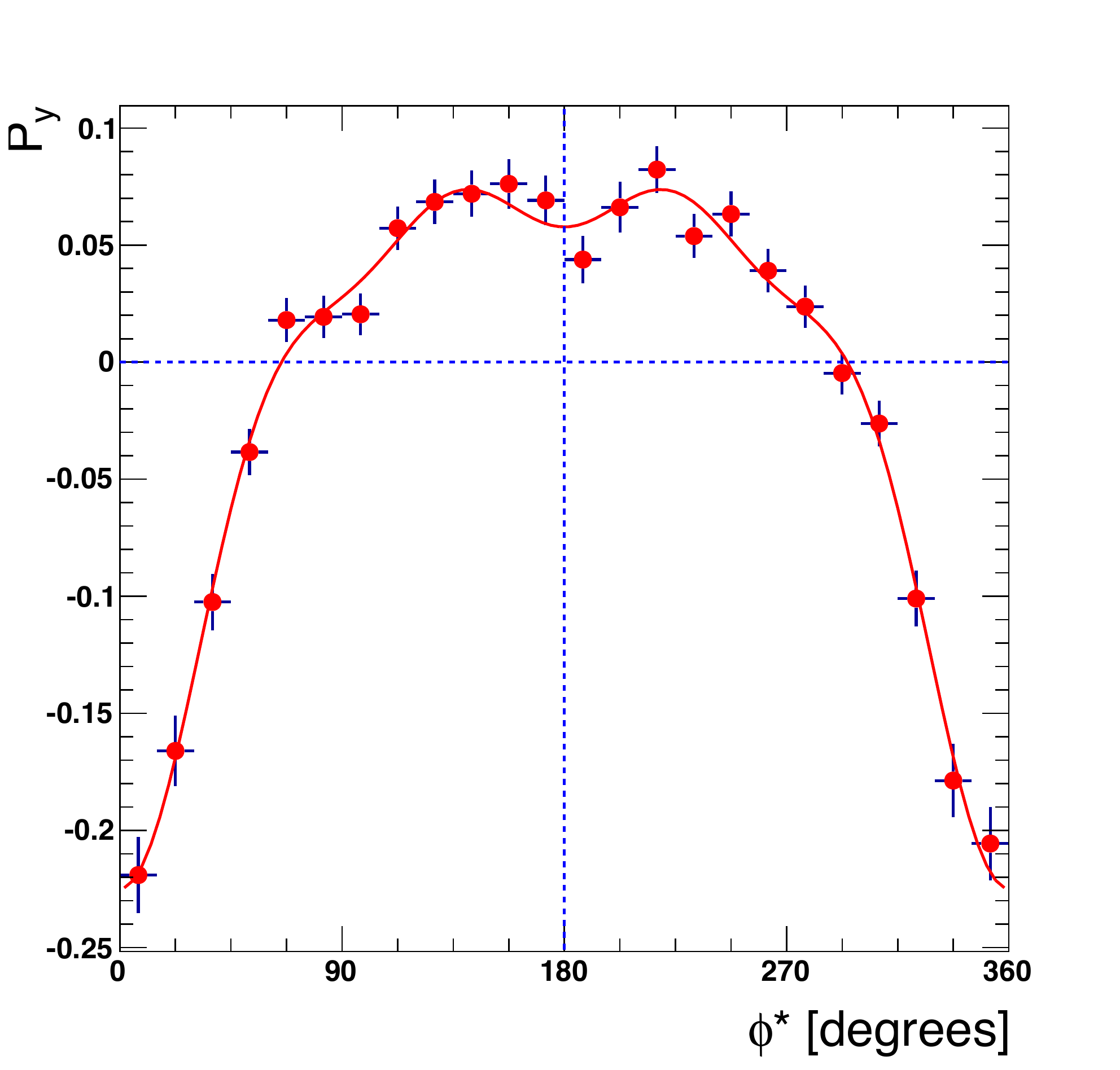}

\includegraphics[width=0.45\textwidth]{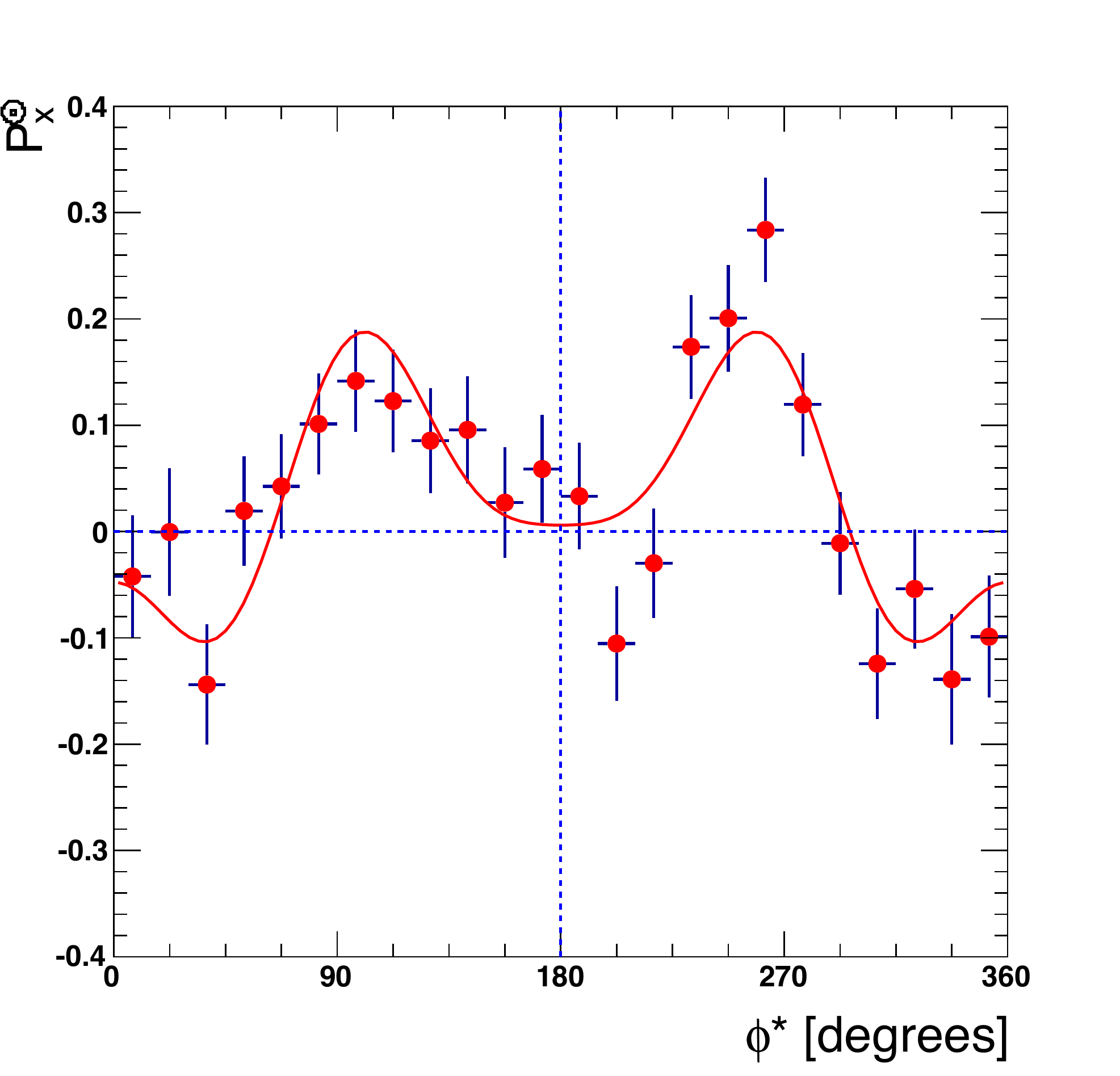}%
\includegraphics[width=0.45\textwidth]{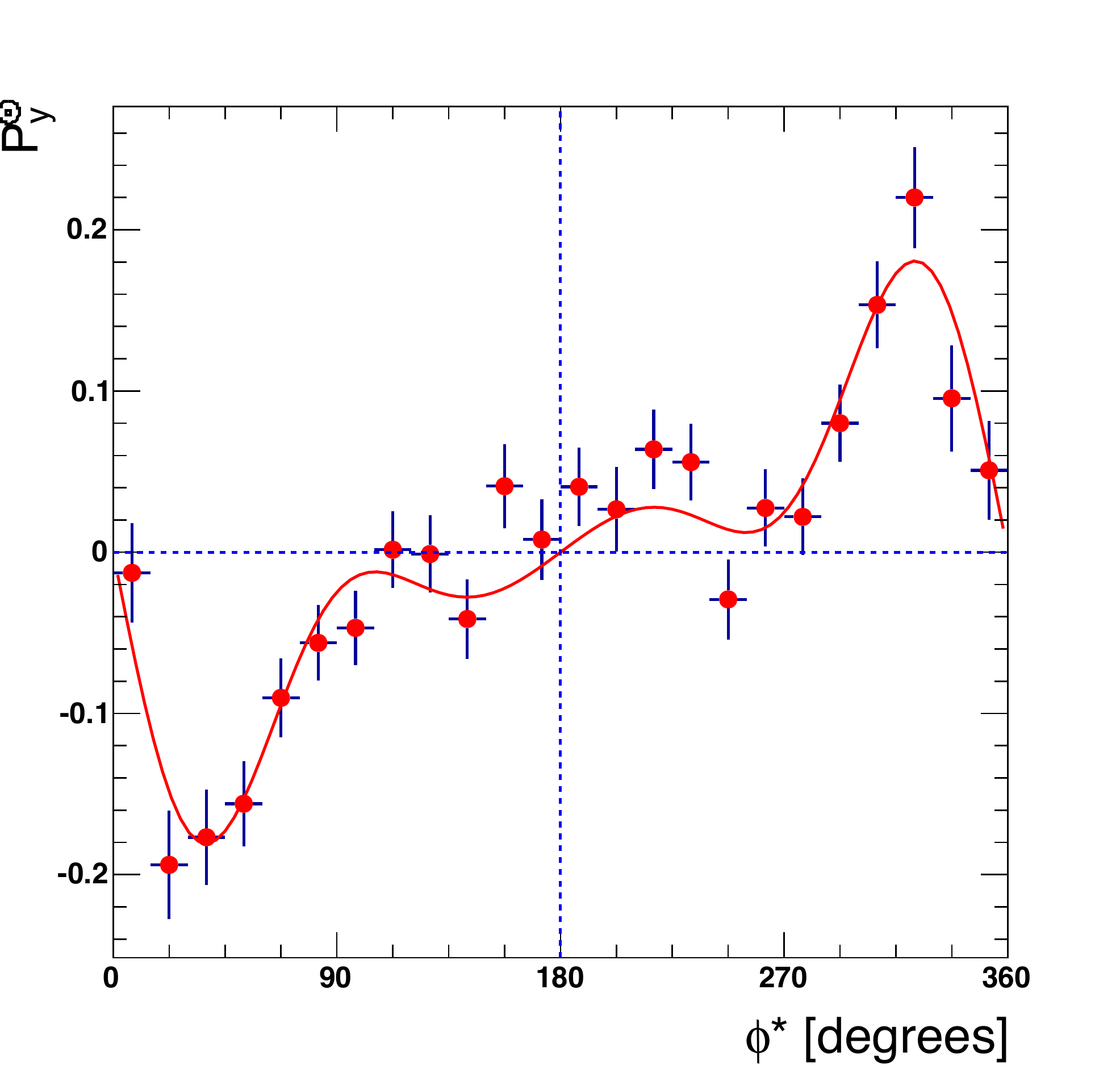}
\caption{Preliminary CLAS results for the polarization observables
  $P_x$, $P_y$, $P^\odot_x$, and $P^\odot_y$ in the
  $\gamma p \to p \pi^+\pi^-$ reaction for
  $1.65~\text{GeV} < W < 1.75~\text{GeV}$. The angle $\phi*$ is the
  $\pi^+$ azimuthal angle in the rest frame of the $\pi^+\pi^-$ system
  with the $z$ direction along the total momentum of the $\pi^+\pi^-$
  system.  The data are fitted with even and odd low-order Fourier
  series according to the symmetry properties of the observables
(red curves). Figure from A.~Net (University of South
  Carolina).}
\label{fig6}
\end{figure}

Many nucleon resonances in the mass region above 1.6~GeV decay
predominantly through either $\pi\Delta$ or $\rho N$ intermediate
states into $\pi\pi N$ final states.  Double-pion photoproduction
allows the study of resonances that have no significant coupling to
the $\pi N$ channel with great potential to observe previously
unobserved states.  Many polarization observables are accessible in
two-pion photoproduction off the nucleon \cite{Roberts:2005mn}.  The
CLAS Collaboration was first to study the beam-helicity asymmetry
$I^\odot$ for the two-pion-photoproduction reaction.  The measurements
covered energies between $W = 1.35$ and 2.30 GeV
\cite{Strauch:2005cs}.  The FROST group is working on the
determination of twelve different polarization observables from
$\gamma p \to p \pi^+\pi^-$ data.  As example, preliminary results are
shown in Fig.~\ref{fig6} for observables that are
accessible in measurements with unpolarized ($P_x$ and $P_y$) and
circularly-polarized ($P^\odot_x$ and $P^\odot_y$) photons off
transversally-polarized protons.  The data will strongly constrain
coupled-channel analyses.

Data have predominantly been taken off proton targets, and the new data
from the FROST program will expand this data base over a large range
of energy with many observables for polarized proton reactions.  In
constrast, data off neutrons are extremely sparse.  However,
measurements with both proton and neutron targets are needed to
completely specify the amplitude of the reaction.

The CLAS collaboration has taken production data with circularly- and
linearly-polarized photons off a polarized solid deuterium-hydride
target (HDice) \cite{Sandorfi:2013gra,PR06101} up to center-of-mass
energies of $W \approx 2.3$~GeV.  The run conditions
were optimized for polarized neutron reactions.  The ongoing analyses
of this run include single- and double-pion photoproduction and
hyperon photoproduction off the bound neutron.

\section{Conclusion}

During the past years, knowledge of the baryon spectrum has 
increased greatly, as reflected in recent $N$ and $\Delta$ resonance
updates in the Review of Particle Physics \cite{Agashe:2014kda}.  New
polarized photoproduction data from CLAS off polarized and
unpolarized proton and neutron targets are both under analysis and
becoming 
available.  These data will contribute to complete or nearly-complete
experiments, and will challenge previously poorly constrained
models. It is very likely that these data will have a tremendous impact on
the understanding of baryon resonances and may provide evidence for
new states found in coupled-channel analyses.

The quasi-real photon tagger for CLAS12 will allow to expand these
photoproduction studies to higher beam energies with new experiments
after the energy upgrade of JLab \cite{Glazier:2010zz}.


\begin{thebibliography}{10}

\bibitem{Crede:2013kia}
V.~Cred\'e and W.~Roberts, {Progress towards understanding baryon resonances},
  Rept. Prog. Phys. \textbf{76} 076301 (2013).

\bibitem{Edwards:2011jj}
R.~G. Edwards, J.~J. Dudek, D.~G. Richards, and S.~J. Wallace, {Excited state
  baryon spectroscopy from lattice QCD}, Phys.~Rev. \textbf{D84} 074508 (2011).

\bibitem{Chiang:1996em}
W.-T. Chiang and F.~Tabakin, {Completeness rules for spin observables in
  pseudoscalar meson photoproduction}, Phys.~Rev. \textbf{C55} 2054 (1997).

\bibitem{Roberts:2005mn}
W.~Roberts and T.~Oed, {Polarization observables for two-pion production off
  the nucleon}, Phys.Rev. \textbf{C71} 055201 (2005).

\bibitem{Strauch:2015noa}
S.~Strauch, {Baryon spectroscopy with polarization observables from CLAS}, in
  \emph{{12th Conference on the Intersections of Particle and Nuclear Physics
  (CIPANP 2015) Vail, Colorado, USA, May 19-24, 2015}} (2015).

\bibitem{Sober:2000we}
D.~Sober \emph{et~al.}, {The bremsstrahlung tagged photon beam in Hall B at
  JLab}, Nucl.~Instrum.~Meth.~A \textbf{440} 263 (2000).

\bibitem{Keith201227}
C.~Keith, J.~Brock, C.~Carlin, S.~Comer, D.~Kashy, J.~McAndrew, D.~Meekins,
  E.~Pasyuk, J.~Pierce, and M.~Seely, {The Jefferson Lab frozen spin target},
  Nucl.~Instrum.~Meth.~A \textbf{684} 27 (2012).

\bibitem{Sandorfi:2013gra}
A.~M. Sandorfi, {Unravelling the excitation spectrum of the nucleon}, J. Phys.
  Conf. Ser. \textbf{424} 012001 (2013).

\bibitem{PR06101}
{Jefferson Lab Experiment E06-101, ``N* Resonances in Pseudoscalar-meson
  photo-production from Polarized Neutrons in $\vec H \cdot \vec D$ and a
  complete determination of the $\gamma n \to K^0\Lambda$ amplitude'', F.~Klein
  and A.M.~Sandorfi, spkespersons}.

\bibitem{Mecking:2003zu}
B.~A. Mecking \emph{et~al.}, {The CEBAF Large Acceptance Spectrometer (CLAS)},
  Nucl.~Instrum.~Meth. \textbf{A503} 513 (2003).

\bibitem{Dugger:2013crn}
M.~Dugger \emph{et~al.}, {Beam asymmetry for $\pi^+$ and $\pi^0$
  photoproduction on the proton for photon energies from 1.102 to 1.862 GeV},
  Phys.~Rev. C \textbf{88} 065203 (2013).

\bibitem{Workman:2012jf}
R.~L. Workman, M.~W. Paris, W.~J. Briscoe, and I.~I. Strakovsky, {Unified
  Chew-Mandelstam SAID analysis of pion photoproduction data}, Phys.~Rev.
  \textbf{C86} 015202 (2012).

\bibitem{McNabb:2003nf}
J.~W.~C. McNabb \emph{et~al.}, {Hyperon photoproduction in the nucleon
  resonance region}, Phys. Rev. \textbf{C69} 042201 (2004).

\bibitem{McCracken:2009ra}
M.~E. McCracken \emph{et~al.}, {Differential cross section and recoil
  polarization measurements for the $\gamma p \to K^+ \Lambda$ reaction using
  CLAS at Jefferson Lab}, Phys. Rev. \textbf{C81} 025201 (2010).

\bibitem{Bradford:2006ba}
R.~K. Bradford \emph{et~al.}, {First measurement of beam-recoil observables
  $C_x$ and $C_z$ in hyperon photoproduction}, Phys. Rev. \textbf{C75} 035205
  (2007).

\bibitem{Sarantsev:2005tg}
A.~V. Sarantsev, V.~A. Nikonov, A.~V. Anisovich, E.~Klempt, and U.~Thoma,
  {Decays of baryon resonances into $\Lambda K^+$, $\Sigma^0 K^+$ and $\Sigma^+
  K^0$}, Eur. Phys. J. \textbf{A25} 441 (2005).

\bibitem{Nikonov:2007br}
V.~A. Nikonov, A.~V. Anisovich, E.~Klempt, A.~V. Sarantsev, and U.~Thoma,
  {Further evidence for $N(1900)P_{13}$ from photoproduction of hyperons},
  Phys. Lett. \textbf{B662} 245 (2008).

\bibitem{PR02112}
Jefferson Lab Experiment E02-112, ``Search for Missing Nucleon Resonances in
  Hyperon Photoproduction'', P.~Eugenio, F.~Klein, and L.~Todor, spokespersons.

\bibitem{PR03105}
Jefferson Lab Experiment E03-105, ``Pion Photoproduction from a Polarized
  Target'', N.~Benmouna, W.~Briscoe, G.~O'Rielly, I.~Strakovsky, S.~Strauch,
  spokespersons.

\bibitem{PR04102}
Jefferson Lab experiment E04-102, ``Helicity Structure of Pion
  Photoproduction'', D.~Crabb, M.~Khandaker, and D.~Sober, spokespersons.

\bibitem{PR05012}
Jefferson Lab experiment E05-012, ``Measurement of polarization observables in
  $\eta$-photoproduction with CLAS'', M.~Dugger and E.~Pasyuk, spokespersons.

\bibitem{PR06013}
Jefferson Lab Experiment E06-013, ``Measurement of $\pi^+\pi^-$ Photoproduction
  in Double-Polarization Experiments using CLAS'', M.~Bellis, V.~Cred\'{e},
  S.~Strauch, spokespersons.

\bibitem{Strauch:2015zob}
S.~Strauch \emph{et~al.}, {First Measurement of the Polarization Observable $E$
  in the $\vec p(\vec \gamma,\pi^+)n$ Reaction up to 2.25 GeV}, Phys. Lett.
  \textbf{B750} 53 (2015).

\bibitem{Anisovich:2015gia}
A. V. Anisovich, V. Burkert, E. Klempt, V. A. Nikonov, E. Pasyuk, A. V.
  Sarantsev, S. Strauch, and U. Thoma, Existence of $\Delta(2200)7/2^-$
  precludes chiral symmetry restoration at high mass, arXiv:1503.05774
  [nucl-ex].

\bibitem{Ronchen:2014cna}
D.~R{\"o}nchen, M.~D{\"o}ring, F.~Huang, H.~Haberzettl, J.~Haidenbauer,
  C.~Hanhart, S.~Krewald, U.~G. Mei{\ss}ner, and K.~Nakayama, {Photocouplings
  at the Pole from Pion Photoproduction}, Eur. Phys. J. \textbf{A50} 101
  (2014).

\bibitem{Anisovich:2011fc}
A.~V. Anisovich, R.~Beck, E.~Klempt, V.~A. Nikonov, A.~V. Sarantsev, and
  U.~Thoma, {Properties of baryon resonances from a multichannel partial wave
  analysis}, Eur. Phys. J. \textbf{A48} 15 (2012).

\bibitem{Senderovich:2015lek}
I.~Senderovich \emph{et~al.}, {First measurement of the helicity asymmetry $E$
  in $\eta$ photoproduction on the proton}, Phys. Lett. \textbf{B755} 64
  (2016).

\bibitem{Drechsel:2007if}
D.~Drechsel, S.~Kamalov, and L.~Tiator, {Unitary Isobar Model - MAID2007}, Eur.
  Phys. J. \textbf{A34} 69 (2007).

\bibitem{WalfordRPR}
T. Corthals, J. Ryckebusch, and T. Van Cauteren, Phys. Rev. {\bf C73} (2006)
  045207 and T.~Vrancx, personal communication.

\bibitem{WalfordKM}
F. X. Lee, T. Mart, C. Bennhold, H. Haberzettl, and L. E. Wright, Nucl. Phys.
  {\bf A695}, 237 (2001), ``Kaon-MAID". Available from
  http://www.kph.uni-mainz.de/MAID/kaon/kaonmaid.html.

\bibitem{Strauch:2005cs}
S.~Strauch \emph{et~al.}, {Beam-helicity asymmetries in double-charged-pion
  photoproduction on the proton}, Phys. Rev. Lett. \textbf{95} 162003 (2005).

\bibitem{Agashe:2014kda}
K.~A. Olive \emph{et~al.}, {Review of Particle Physics}, Chin. Phys.
  \textbf{C38} 090001 (2014).

\bibitem{Glazier:2010zz}
D.~I. Glazier, {A quasi-real photon tagger for CLAS12}, Nucl. Phys. Proc.
  Suppl. \textbf{207-208} 204 (2010).

\end{thebibliography}
\end{document}